\documentclass[preprint,pra,showpacs,nofootinbib]{revtex4}
\usepackage{mathrsfs}
\usepackage{float,epsfig}
\usepackage{dcolumn}
\usepackage{bm}
\usepackage{graphicx}
\usepackage{dcolumn}
\usepackage{bm}
\usepackage{amsmath,amssymb,amsthm}
\newcommand{\bea}{\begin{eqnarray}}
\newcommand{\eea}{\end{eqnarray}}
\newcommand{\beq}{\begin{equation}}
\newcommand{\eeq}{\end{equation}}

\begin{document}
\title{Entanglement generation outside a Schwarzschild black hole and the Hawking effect}
\author{  Jiawei Hu$^{1}$ and Hongwei Yu$^{1,2}$\footnote{Corresponding author}  }
\affiliation{$^1$ Institute of Physics and Key Laboratory of Low
Dimensional Quantum Structures and Quantum
Control of Ministry of Education,\\
Hunan Normal University, Changsha, Hunan 410081, China \\
$^2$ Center for Nonlinear Science and Department of Physics, Ningbo
University,  Ningbo, Zhejiang 315211, China}

\begin{abstract}

We examine the Hawking effect by studying the asymptotic
entanglement of two mutually independent two-level atoms  placed at
a fixed radial distance outside a Schwarzschild black hole in the
framework of open quantum systems. We treat the two-atom system as
an open quantum system in a bath of fluctuating  quantized massless
scalar fields in vacuum and calculate the concurrence, a measurement
of entanglement, of the equilibrium state of the system at large
times, for the Unruh, Hartle-Hawking and Boulware vacua
respectively. We find, for all three vacuum cases,  that the atoms
turn out to be entangled even if they are initially in a separable
state as long as the system is not placed right at the even horizon.
Remarkably, only in the Unruh vacuum, will the asymptotic
entanglement be affected by the backscattering of the thermal
radiation off the space-time curvature. The effect of the back
scatterings on the asymptotic entanglement cancels in the
Hartle-Hawking vacuum case.

\end{abstract}

\maketitle

\section{Introduction}

Classically, black holes are described as massive objects with such
a strong gravitational field that even light cannot escape from
them. However, Hawking finds, in the framework of quantum field
theory in curved space-time, that a black hole is not completely
black, but emits thermal radiation with a black body spectrum due to
quantum effects~\cite{hk}. Ever since this surprising  discovery,
the Hawking effect has attracted widespread interest in the physics
community and extensive works have been done trying to understand it
in various different physical contexts (See for example,
Ref.~\cite{hk,Gibbons,Parikh,SC,Robinson,hkur,yu1,yu2,yu3}).

In this paper, we will try to examine the Hawking effect in terms of
the entanglement generation in the framework of open quantum
systems. The open system we are going to study consists of two
mutually independent static two-level atoms subjected to a bath of
fluctuating  quantized massless  scalar fields in vacuum outside a
Schwarzschild black hole. We will analyze the time evolution of the
density matrix describing the system using the well-known techniques
in the theory of open quantum systems.  Let us note that the reduced
dynamics of a single static detector (a two-level atom) placed
outside a Schwarzschild black hole interacting with quantized
massless scalar fields in the Unruh, Hartle-Hawking and Boulware
vacua has been recently investigated~\cite{yu3} and it has been
found that in both the Unruh and Hartle-Hawking vacua, the detector
will spontaneously excite with a nonvanishing probability  as if
there were thermal radiation at the Hawking temperature. An another
way to look at the issue is to study the equilibrium state of the
detector. In this regard, it has been shown that the detector is
asymptotically driven to a thermal state at the Hawking temperature
in the spatial asymptotic region, regardless of its initial
state~\cite{ZY}. This approach has also been  applied to reproduce
both the Unruh effect and the Gibbons-Hawking effect, in
Ref.~\cite{Benatti1} and~\cite{yu4}, respectively.

If the single atom is replaced by two independent two-level atoms,
the situation becomes physically more interesting. A study of the
reduced  dynamics of the two-atom system may reveal whether the
asymptotic equilibrium state of the system will be entangled or not.
It is known that an environment usually leads to decoherence and
noise, which may cause entanglement that might have been created
before to disappear. However, in certain circumstances, the
environment may enhance entanglement rather than destroying
it~\cite{pr1,pr2,pr3,pr4,pr5,pr6}. The reason is that an external
environment can also provide an indirect interaction between
otherwise totally uncoupled subsystems through correlations that
exist.  It is of interest to see whether a bath of fluctuating
vacuum scalar fields outside a Schwarzschild black hole can provide
such an indirect coupling to enhance the entanglement. This is what
we are going to pursue in the present paper. We will, in the hope of
gaining new understanding of the Hawking effect from a different
perspective,  examine the asymptotic entanglement of the two-atom
system in a bath of massless scalar fields in the
Unruh~\cite{Unruh}, Hartle-Hawking~\cite{Hartle-Hawking}, and
Boulware~\cite{Boulware} vacua outside a black hole.  Let us note
here that a similar issue related to the Unruh effect has already
been worked out by Benatti and Floreanini, in which they have
studied a uniformly accelerating two-atom system~\cite{Benatti1} and
found that the two initially separable atoms will become entangled,
just as if they were immersed in a thermal bath at the Unruh
temperature.


\section{The Master Equation}

The system we study consists of  two mutually independent static
two-level atoms of vanishing separation in interaction with a bath
of fluctuating quantum scalar fields in vacuum outside a
Schwarzschild black hole which is described by the following metric
\begin{equation}
ds^2=\bigg(1-\frac{2M}{r}\bigg)dt^2-\frac{dr^2}{1-2M/r}-r^2(d\theta^2+\sin^2\theta{d\phi^2})\;.
\end{equation}
 Although the atoms are mutually independent, the fluctuating
vacuum fields with which the atoms are coupled may provide an
indirect interaction and therefore a means to generate entanglement
between them. In this paper, we are particularly concerned with the
issue of  whether the atoms can be entangled when system reaches an
equilibrium state as time becomes large. Generically, the total
Hamiltonian of the system takes the form
\begin{equation}\label{H}
 H=H_s+H_\phi+\lambda\;H'\;.
\end{equation}
Here $H_s$ is the Hamiltonian of the two atoms,
 \begin{equation}
H_s=H_s^{(1)}+H_s^{(2)},\ \ H_s^{(\alpha)}={\omega\over 2}\,
n_i\,\sigma_i^{(\alpha)}, \quad(\alpha=1,2), \ \
 \end{equation}
where $\sigma_i^{(1)}=\sigma_i \otimes \sigma_0,\ \
\sigma_i^{(2)}=\sigma_0 \otimes \sigma_i$,  $\sigma_i\; (i=1,2,3)$
are the Pauli matrices, $\sigma_0$ the $2\times2$ unit matrix,
$\mathbf{n} =(n_1,n_2,n_3)$ a unit vector, and $\omega_0$ the energy
level spacing. $H_\phi$ is the Hamiltonian of free massless scalar
fields. The Hamiltonian $H'$ that describes the interaction between
the two atoms with the external scalar fields is given by
\begin{equation}
H'=\sum_{\mu=0}^{3}\,[(\sigma_{\mu} \otimes \sigma_0)\Phi_{\mu}(t,
{\bf x}_1)+(\sigma_{0} \otimes \sigma_{\mu})\Phi_{\mu}(t, {\bf
x}_2)\,]\;.
\end{equation}
We assume that the scalar fields can be expanded as
\begin{equation}
\Phi_{\mu}(x)=\sum^N_{a=1}\,[\chi_\mu^a\phi^{(-)}(x) +
(\chi_\mu^a)^*\phi^{(+)}(x)]\;,
\end{equation}
where $\phi^{(\pm)}(x)$ are positive and negative energy field
operators of the massless scalar field, and $\chi_\mu^a$ are the
corresponding complex coefficients.

At the beginning, the whole system is characterized by the total density matrix
$\rho_{tot}=\rho(0) \otimes |0\rangle\langle0|$, in which $\rho(0)$ is the initial reduced density matrix of the two-atom system, and $|0\rangle$ is the vacuum state of field $\Phi(x)$. In the frame of the two-atom system, the evolution in the proper time $\tau $ of the total density matrix $\rho_{tot}$ satisfies
\begin{equation}
\frac{\partial\rho_{tot}(\tau)}{\partial\tau}=-iL_H[\rho_{tot}(\tau)]\;,
\label{evolution}
\end{equation}
where the symbol $L_H$ represents the Liouville operator
associated with $H$
\begin{equation}
L_H[S]\equiv[H,S]\;.
\end{equation}
We assume that the interaction between the atoms and the field is
weak, i.e., the coupling constant $\lambda$ in (\ref{H}) is small.
In the limit of weak coupling, the evolution of the reduced density
matrix $\rho(\tau)$ can be written in the Kossakowski-Lindblad
form~\cite{Lindblad, Benatti1, Benatti2, pr5}
\begin{equation}
{\partial\rho(\tau)\over \partial \tau}= -i \big[H_{\rm eff},\,
\rho(\tau)\big]
 + {\cal L}[\rho(\tau)]\ ,
\label{master}
\end{equation}
with
\begin{equation}
H_{\rm eff}=H_s-\frac{i}{2}\sum_{\alpha,\beta=1}^2\sum_{i,j=1}^3
H_{ij}\ \sigma_i^{(\alpha)}\,\sigma_j^{(\beta)}\ ,
\label{}
\end{equation}
and
\begin{equation}
{\cal L}[\rho]={1\over2} \sum_{\alpha,\beta=1}^2\sum_{i,j=1}^3
C_{ij}\big[2\,
\sigma_j^{(\beta)}\rho\,\sigma_i^{(\alpha)}
-\sigma_i^{(\alpha)}\sigma_j^{(\beta)}\, \rho
-\rho\,\sigma_i^{(\alpha)}\sigma_j^{(\beta)}\big]\ .
\label{lindblad}
\end{equation}
The coefficients of the matrix $C_{ij}$ and $H_{ij}$ are determined
by the Fourier and Hilbert transforms of the field correlation
functions
\begin{equation}\label{correlation}
G_{ij}(x-y)=\langle0|\Phi_i(x)\Phi_j(y)|0 \rangle\;,
\end{equation}
which are defined as follows
\begin{equation}
{\cal G}_{ij}(\lambda)=\int_{-\infty}^{\infty} d\tau \,
e^{i{\lambda}\tau}\, G_{ij}(\tau)\; , \label{fourierG}
\end{equation}
\begin{equation}
{\cal K}_{ij}(\lambda)=\int_{-\infty}^{\infty} d\tau \,
{\rm sign}(\tau)\, e^{i{\lambda}\tau}\, G_{ij}(\tau)=
\frac{P}{\pi i}\int_{-\infty}^{\infty} d\omega\ \frac{ {\cal
G}_{ij}(\omega) }{\omega-\lambda} \;,
\end{equation}
in which $P$ denotes principal value. It can be shown that the
Kossakowski matrix $C_{ij}$ can be written
explicitly as
\begin{equation}
C_{ij}=\sum_{\xi=+,-,0}\sum_{k,l=1}^3 {\cal
G}_{kl}(\xi\omega)\, \psi_{ki}^{(\xi)}\,
\psi_{lj}^{(-\xi)}\;,
\end{equation}
where
\begin{equation}
\psi_{ij}^{(0)}=n_i\, n_j\ ,\qquad \psi_{ij}^{(\pm)}={1\over
2}\big(\delta_{ij} - n_i\, n_j\pm i\epsilon_{ijk} n_k\big)\ .
\label{psij}
\end{equation} Similarly, the coefficients of $H_{ij}$
can be obtained by replacing ${\cal
G}_{kl}(\xi\omega)$ with ${\cal
K}_{kl}(\xi\omega)$ in the above equations.

If we assume that the field components $\Phi_i(x)$ are independent,
or the coefficients $\chi^a_\mu$ satisfy
 \begin{equation}
\sum^N_{a=1}\,\chi^a_\mu(\chi^a_\nu)^*=\delta_{\mu\nu}\;,
\end{equation}
the field correlation functions in (\ref{correlation}) are diagonal such that
\begin{equation}
G_{ij}(x-y)=\delta_{ij}G(x-y)\;,
\end{equation}
in which $G(x-y)$ is the standard Wightman function, and the Kossakowski matrix $C_{ij}$
can be written as~\cite{Benatti1}
\begin{equation}
C_{ij}=A\;\delta_{ij}-iB\; \epsilon_{ijk}n_k+C\;n_in_j\; ,
\label{cij}
\end{equation}
where
\begin{equation}
A=\frac{1}{2}[{\cal {G}}(\omega_0)+{\cal{G}}(-\omega_0)]\;,\;~~~
B=\frac{1}{2}[{\cal {G}}(\omega_0)-{\cal{G}}(-\omega_0)]\;,~~~~
C={\cal{G}}(0)-A\;.\ \label{abc}
\end{equation}

\section{the asymptotic entanglement}

In order to figure out whether the entanglement will be generated
between the atoms and whether it can persist asymptotically through
the analysis of the master equation governing the density matrix, we
need to calculate the coefficients $C_{ij}$ which are determined by
the field correlation function in the vacuum state. However, when a
vacuum state is concerned in a curved space-time, a delicate issue
then arises as to how the vacuum state of the quantum fields is
determined. Normally, a vacuum state is associated with
non-occupation of positive frequency modes. However, the positive
frequency of field modes is defined with respect to the time
coordinate. Therefore, to define positive frequency, one has to
first specify a definition of time. In a spherically symmetric black
hole background, three different vacuum states, i.e., the Unruh~\cite{Unruh}, Hartle-Hawking~\cite{Hartle-Hawking}, and Boulware~\cite{Boulware} vacuum states, have been defined, each corresponding to a different choice of time coordinate. In the next section, we examine the asymptotic entanglement for the atoms in all vacua.

 \subsection{The Unruh vacuum}

Let us begin our discussion with the Unruh vacuum which is supposed
to be the vacuum state best approximating the state  following the
gravitational collapse of a massive body to a black hole. In order
to analyze the asymptotic state of the system, we will deal with the
master equation~(\ref{master}) and see how the reduced density
matrix of the two-atom system $\rho$ evolves.

Generally, we can write the reduced density matrix of the two-atom system in the form of
\begin{equation}\label{rho-1}
\rho(\tau)={1\over4}\bigg[\sigma_0\otimes\sigma_0+\rho_{0i}(\tau)\;\sigma_0\otimes\sigma_i
+\rho_{i0}(\tau)\sigma_i\otimes\sigma_0+\rho_{ij}(\tau)\sigma_i\otimes\sigma_j\bigg]\;,
\end{equation}
which is normalized as ${\rm {Tr}}(\rho)=1$ with ${\rm
{det}}(\rho)\geq0\;.$ Here we are interested in the final
equilibrium density matrix $\rho^\infty$, which does not change with
time, i.e., $\partial_\tau{\rho^\infty}=0$.  Let us note that the
unitary term involving  Hamiltonian $H_{\rm eff}$ in the master
equation can be ignored since it does not give rise to the
entanglement phenomena~\cite{Benatti1} and we only need to examine
the effects produced by the dissipative term ${\cal L}[\rho]$ in
(\ref{master}). As a result, the equilibrium condition becomes
${\cal L}[\rho^\infty]=0$. After some direct calculations, one obtains
the components of $\rho^\infty$~\cite{Benatti1}
 \begin{eqnarray}
&&\rho_{0i}^\infty=\rho_{i0}^\infty=-{R\over 3+R^2}(\tau_*+3)n_i\;, \nonumber \\
&&\rho_{ij}^\infty={1\over3+R^2}[(\tau_*-R^2)\delta_{ij}+R^2(\tau_*+3)n_in_j]\,,
 \end{eqnarray}
where $R=B/A$, $\tau_*$ is the trace of the density matrix
$\tau_*=\Sigma_{i=1}^3\rho_{ii}(\tau)$, which is actually a constant
of motion, and the positivity of $\rho(0)$ requires
$-3\leq\tau_*\leq1$.

In order to determine whether the final equilibrium state is
entangled or not, we take the concurrence as a measurement of the
entanglement, which is defined as $ {\cal C}[\rho]
=\max\{0,\lambda_1-\lambda_2 -\lambda_3-\lambda_4\}\;$, where
 $\lambda_{\mu}\;(\mu=1,2,3,4\;)$ are the square roots of the
non-negative eigenvalues of the matrix
$\rho(\sigma_2\otimes\sigma_2)\rho^{T}(\sigma_2\otimes\sigma_2)$ in
decreasing order, and $T$ stands for transposition. The value of ${\cal C}[\rho]$ ranges from 0, for separable states, to 1, for maximally entangled states. For the current case, the concurrence is
\begin{equation}
{\cal
C}(\rho^\infty)=\max\Bigg\{{(3-R^2)\over2(3+R^2)}\bigg[{5R^2-3\over3-R^2}-\tau_*\bigg],0\Bigg\}\;,
\end{equation}
which is non-zero provided $\tau_*$ obeys
\begin{equation}\label{ent-con}
\tau_* < {5R^2-3 \over 3-R^2}\;.
\end{equation}
This result implies that as long as the condition~(\ref{ent-con}) is
satisfied, the equilibrium state will turn out to be entangled, even
if the initial state is separable.

In order to make our discussion more concise, let us come to a
simple example. The initial state of the system is taken to be a
separable state provided by the direct product of two pure states:
\begin{equation}
\rho(0)=\rho_n\otimes\rho_m\ , \qquad \rho_n={1\over2}\Big(1 + \vec
n\cdot\vec\sigma\Big)\ , \quad \rho_m={1\over2}\Big(1 + \vec
m\cdot\vec\sigma\Big)\ ,
\end{equation}
where $\vec n$ and $\vec m$ are two unit vectors. Here $\tau_*=\vec
n\cdot\vec m$, so the asymptotic entanglement is maximized when
$\vec n=-\vec m$:
\begin{equation}\label{cr}
{\cal C}[\rho^\infty]={2R^2\over 3+R^2}\;.
\end{equation}
Eq.~(\ref{cr}) shows that, in our simple case, the concurrence
increases with $R$ monotonically when $R$ ranges from 0 to 1, and it
reaches its maximum ${\cal C}[\rho^\infty]=1/2$ for $R=1$. So, we
need only evaluate $R$ in order to analyze the entanglement of the
system.

Now let us begin to compute $R$ in the Unruh vacuum case. The
Wightman function for massless scalar fields in the Unruh vacuum
is given by~\cite{wightman1,wightman2,wightman3}
 \begin{equation}
{G^+}(x,x')=
\sum_{ml}\int^{\infty}_{-\infty}\frac{e^{-i\omega\Delta{t}}}{4
\pi\omega}|\,Y_{lm}(\theta,\phi)\,|^2\bigg[\frac{|\,\overrightarrow{R_l}(\omega,r)\,|^2}{1-e^{-2\pi\omega/\kappa}}
+\theta(\omega)|\,\overleftarrow{R_l}(\omega,r)\,|^2\bigg]d\omega\;,
 \end{equation}
where $\kappa=1/4M$ is the surface gravity of the black hole. Its
Fourier transform is
\begin{eqnarray}\label{guf}
{\cal{G}}(\lambda)&=&\int^{\infty}_{-\infty}e^{i{\lambda}\tau}{G^+}(x,x')d\tau\nonumber\\
&=&\frac{1}{8\pi{\lambda}}\sum_{l=0}^{\infty}\bigg[\theta({\lambda}\sqrt{g_{00}})(1+2l)|\,\overleftarrow{R_l}({\lambda}\sqrt{g_{00}},r)\,|^2
+\frac{(1+2l)|\,\overrightarrow{R_l}({\lambda}\sqrt{g_{00}},r)\,|^2}{1-e^{-2\pi{\lambda}\sqrt{g_{00}}/\kappa}}\bigg]\;,
\end{eqnarray}
where we have used the relation
\begin{equation}
 \sum^l_{m=-l}|\,Y_{lm}(\,\theta,\phi\,)\,|^2= {2l+1 \over
 4\pi}\;,
 \end{equation}
and here $\kappa_r$ is defined as $\kappa/\sqrt{g_{00}}\;$. The
above Fourier transform, which is needed in our computation of $R$,
is hard to evaluate, since we do not know the exact form of the
radial functions $\overrightarrow{R_l}(\omega,r)$ and
$\overleftarrow{R_l}(\omega,r)$. Here, we choose to compute it both
close to the event horizon and at infinity. For this purpose,  let
us recall that the radial functions have the following properties in
asymptotic regions~\cite{wightman3}:
\begin{equation} \label{asymp1}
\sum_{l=0}^\infty\,(2l+1)\,|\overrightarrow{R}_l(\,\omega,r\,)\,|^2\sim\left\{
                    \begin{aligned}
                 &\frac{4\omega^2}{1-\frac{2M}{r}}\;,\;\;\;\quad\quad\quad\quad\quad\quad\quad r\rightarrow2M\;,\cr
                  &\frac{1}{r^2}
\sum_{l=0}^\infty(2l+1)\,|\,{B}_l\,(\omega)\,|^2\;,\quad\;r\rightarrow\infty
                  \;,
                          \end{aligned} \right.
\end{equation}
\begin{equation} \label{asymp2}
\sum_{l=0}^\infty\,(2l+1)\,|\overleftarrow{R}_l(\,\omega,r\,)\,|^2\sim\left\{
                    \begin{aligned}
                 &\frac{1}{4M^2}\sum_{l=0}^\infty(2l+1)\,|\,{B}_l\,(\omega)\,|^2,\quad\;r\rightarrow2M\;,\cr
                  &\frac{4\omega^2}{1-\frac{2M}{r}},\;\;\;\;\quad\quad\quad\quad\quad\quad\quad\quad r\rightarrow\infty
                  \;.\cr
                          \end{aligned} \right.
\end{equation}
Inserting Eq.~(\ref{guf}) into Eq~(\ref{abc}), and using
Eq.~(\ref{asymp1}) and ~(\ref{asymp2}), one finds that
\begin{eqnarray}
r\rightarrow2M:\left\{ \begin{aligned}
 &A\approx
 \frac{\omega_0}{4\pi}\;[1+g_{00}\;f(\omega_0\sqrt{g_{00}},2M)
 +\frac{2}{e^{2\pi\omega_0/\kappa_r}-1}]\;,\\
 &B\approx
 \frac{\omega_0}{4\pi}\;[1+g_{00}\;f(\omega_0\sqrt{g_{00}},2M)]\;,
 \end{aligned} \right.
 \end{eqnarray}
 and
\begin{eqnarray}
r\rightarrow\infty:\left\{ \begin{aligned}
 &A\approx
 \frac{\omega_0}{4\pi}\;[1+g_{00}\;f(\omega_0\sqrt{g_{00}},r)+\frac{2}{e^{2\pi\omega_0/\kappa_r}-1}
 g_{00}f(\omega_0\sqrt{g_{00}},r)]\;,\\
 &B\approx
 \frac{\omega_0}{4\pi}\;[1+g_{00}\;f(\omega_0\sqrt{g_{00}},r)]\;,
 \end{aligned} \right.
 \end{eqnarray}
where $f(\omega,r)$ is defined as
  \begin{equation}
f(\omega,r)=\frac{1}{4\,r^2\omega^2}\,\sum_{l=0}^\infty\,(2l+1)\,|\,B_l\,(\,\omega)|^2\;.
 \end{equation}
 Straightforward calculations then yield in the asymptotic regions,
\begin{equation} R={B\over A}=\left\{
\begin{aligned}
 &\frac{1+g_{00}f(\omega_0\sqrt{g_{00}},2M)}{1+g_{00}f(\omega_0\sqrt{g_{00}},2M)+\frac{2}{e^{{2\pi\omega_0}/{\kappa_r}}-1}}\;,&r\rightarrow2M\;,
\\
&\frac{1+g_{00}f(\omega_0\sqrt{g_{00}},r)}{1+g_{00}f(\omega_0\sqrt{g_{00}},r)+\frac{2}{e^{{2\pi\omega_0}/{\kappa_r}}-1}g_{00}f(\omega_0\sqrt{g_{00}},r)}
\;,&r\rightarrow\infty\;.
\end{aligned} \right. \label{aaa}
\end{equation}
We can  see that, in the vicinity of the event horizon, the first
two terms in the denominator are the same as their counter-parts in
the numerator, and the third one is the standard Planckian factor.
At infinity, the Planckian factor is modified by a grey-body factor
$g_{00} f(\omega_0\sqrt{g_{00}},r)$ caused by the backscattering off
the space-time curvature. This suggests that at the horizon, there
is a thermal flux going outwards, which is weakened by the
backscattering off the curvature on its way to infinity. Actually,
it has  been shown that, in the framework of open quantum system,
the spontaneous excitation rate per unit time of a particle detector
from the initial ground state $i$ to the final excited state $f$ is
just~\cite{Benatti1,yu3}
\begin{equation}\label{gammaif}
{\Gamma_{i\rightarrow{f}}}=2(A-B)=2{\cal{G}}(-\omega_0)\;.
\end{equation}
This means that the difference between the denominator and the
numerator of $R$ is proportional to the spontaneous excitation rate
per unit time of a particle detector, i.e., the strength of the
thermal radiation. The larger the difference between $A$ and $B$,
the stronger the thermal radiation, and the less the two-atom system
gets entangled.

Here we note that, using the geometrical optics approximation~\cite{wightman1}, the transmission amplitude $B_l(\omega)$ can be approximated as $B_l(\omega)\sim\theta(\sqrt{27}M\omega-l)$, where $\theta(x)$ is the standard step function, which gives 0 for $x<0$ and 1 for $x>0$. So
$g_{00}f(\omega_0\sqrt{g_{00}},r)$ can be simplified as
 \bea
g_{00}f(\omega_0\sqrt{g_{00}},r)
\approx\frac{27M^2g_{00}}{4r^2}=\frac{27M^2}{4r^2}\bigg(1-{2M\over
r}\bigg)\equiv f(r)\;,
 \eea
and in both the asymptotic regions,
$g_{00}f(\omega_0\sqrt{g_{00}},r)\rightarrow0$. Allowing for this,
we have,  in the vicinity of the event horizon,
\begin{equation}
R=\frac{e^{2\pi\omega_0/\kappa_{r}}-1}{e^{2\pi\omega_0/\kappa_{r}}+1}\;. \label{aaa}
\end{equation}
It is interesting to note that the $R$ we obtain here is exactly the
same as that in the case of a two-atom system immersed in a thermal
bath at the temperature $T=\kappa_r/2\pi=T_H/\sqrt{g_{00}}$ in a
flat space-time, where $T_H=\kappa/2\pi$ is the Hawking
temperature~\cite{hk}. When $r\rightarrow 2M$, the temperature $T$
is divergent, since this effective temperature is a result of both
the thermal flux from the black hole and the Unruh effect due to
that the system is accelerating with respect to the local
free-falling inertial frame so as to maintain at a fixed distance
from the black hole, and the acceleration diverges at the horizon.
In this case, the concurrence ${\cal C}[\rho^\infty]$ approaches
zero, which means that final equilibrium state of the two atom
system, which is very close to the event horizon, will not be
entangled. As the atoms are placed farther, then  the thermal
radiation becomes weaker due to the back scattering off the
space-time curvature and the concurrence grows larger. At the
infinity, the grey-body factor vanishes and $R$ approaches 1, so
that  the concurrence ${\cal C}[\rho^\infty]$ tends to reach its
maximum $1/2$. This result suggests that, in the Unruh vacuum,  no
thermal radiation is felt at the infinity due to the back scattering
of the outgoing thermal radiation off the space-time curvature.  Here
it is clear that in the vicinity of the horizon, the entanglement of
the system is enhanced due to the back scattering, while in the
infinity, the concurrence is smaller than its maximum if we allow
for the thermal radiation emitted from the horizon although it is
back scattered by the space-time curvature.

Compared with the case of a thermal bath in a flat space-time, we
find that the parameters $A$ and $B$ are modified by the grey-body
factor $f(\omega,r)$, which is a function of $M$ and $r$ for a given
energy gap $\omega_0$. A similar result is derived when studying the
entanglement generation in atoms immersed in a thermal bath of
external quantum scalar fields with a reflecting
boundary~\cite{zhang}. This result implies that the back scattering
of vacuum field modes off the space-time curvature of the black hole
manifests  in much the same way as the reflection of the field modes
at the reflecting boundary in a flat space-time.

 \subsection{The Hartle-Hawking vacuum}

Now let us move on to the Hartle-Hawking vacuum. The Wightman
function is now given by~\cite{wightman1,wightman2,wightman3}
\begin{eqnarray}
{G^+}(x,x')=\sum_{ml}\int^{\infty}_{-\infty}\frac{|\,Y_{lm}(\theta,\phi)\,|^2}{4
\pi\omega}\bigg[\frac{e^{-i\omega{\Delta{t}}}}{1-e^{-2\pi\omega/\kappa}}|\,\overrightarrow{R_l}(\omega,r)\,|^2+\frac{
e^{i\omega\Delta{t}}}{e^{2\pi\omega/\kappa}-1}|\,\overleftarrow{R_l}(\omega,r)\,|^2\bigg]d\omega\;,\nonumber\\
\end{eqnarray}
and its Fourier transform is
\begin{eqnarray}
{\cal{G}}(\lambda)&=&\int^{\infty}_{-\infty}e^{i{\lambda}\tau}{G}^+(x,x')d\tau
\nonumber\\&=&\sum_{l=0}^{\infty}\frac{(1+2l)}{8\pi{{\lambda}}}\bigg[
\frac{|\,\overrightarrow{R_l}({\lambda}\sqrt{g_{00}},r)\,|^2}{1-e^{-2\pi{{\lambda}}/\kappa_r}}+
\frac{|\,\overleftarrow{R_l}(-{\lambda}\sqrt{g_{00}},r)\,|^2}{1-e^{-2\pi{{\lambda}}/\kappa_r}}\bigg]\;,
\end{eqnarray}
Similar calculations then lead to
\begin{eqnarray}
r\rightarrow2M:\left\{ \begin{aligned}
 &A\approx
 \frac{\omega_0}{4\pi}\frac{e^{2\pi\omega_0/\kappa_r}+1}{e^{2\pi\omega_0/\kappa_r}-1}\;
 [1+g_{00}\;f(\omega_0\sqrt{g_{00}},2M)]\;,\\
 &B\approx
 \frac{\omega_0}{4\pi}\;[1+g_{00}\;f(\omega_0\sqrt{g_{00}},2M)]\;,
 \end{aligned} \right.
 \end{eqnarray}
\begin{eqnarray}
r\rightarrow\infty:\left\{ \begin{aligned}
 &A\approx
 \frac{\omega_0}{4\pi}\frac{e^{2\pi\omega_0/\kappa_r}+1}{e^{2\pi\omega_0/\kappa_r}-1}\;
 [1+g_{00}\;f(\omega_0\sqrt{g_{00}},r)]\;,\\
 &B\approx
 \frac{\omega_0}{4\pi}\;[1+g_{00}\;f(\omega_0\sqrt{g_{00}},r)]\;,
 \end{aligned} \right.
 \end{eqnarray}
and
\begin{equation} R=\frac{B}{A}=\left\{
\begin{aligned}
 &\frac{1+g_{00}f(\omega_0\sqrt{g_{00}},2M)}{1+g_{00}f(\omega_0\sqrt{g_{00}},2M)
 +\frac{2}{e^{{2\pi\omega_0}/{\kappa_r}}-1}+\frac{2}{e^{{2\pi\omega_0}/{\kappa_r}}-1}g_{00}f(\omega_0\sqrt{g_{00}},2M)}\;,&r\rightarrow2M\;,
\\
&\frac{1+g_{00}f(\omega_0\sqrt{g_{00}},r)}{1+g_{00}f(\omega_0\sqrt{g_{00}},r)
 +\frac{2}{e^{{2\pi\omega_0}/{\kappa_r}}-1}+\frac{2}{e^{{2\pi\omega_0}/{\kappa_r}}-1}g_{00}f(\omega_0\sqrt{g_{00}},r)}
\;,&r\rightarrow\infty\;.
\end{aligned} \right. \label{34}
\end{equation}
Unlike that in the Unruh vacuum case, here $R$ is the same in both
the asymptotic regions. We have two Planckian terms in the
denominator. One is the standard one and the other is a Planckian
factor modified by a grey-body factor caused by the backscattering
off the space-time curvature. This suggests that there are  thermal
radiation outgoing from the horizon and that incoming from infinity,
both of which are weakened by the backscattering off the curvature
on their way. Therefore, the Hartle-Hawking vacuum is actually a
state that describes a black hole in equilibrium with an infinite
sea of blackbody radiation~\cite{wightman3}.

Here we notice that Eq.~(\ref{34}) can be simplified as
\begin{eqnarray}\label{b1b21}
R={B \over
A}=\frac{e^{2\pi\omega_0/\kappa_{r}}-1}{e^{2\pi\omega_0/\kappa_{r}}+1}\;.
\end{eqnarray}
At close to the horizon, i.e., when $r\rightarrow2M$,
$R\rightarrow0$, which means that final state of the two-atom system
is not entangled. When $r\rightarrow \infty$,
$R\rightarrow\frac{e^{2\pi\omega_0/\kappa}-1}{e^{2\pi\omega_0/\kappa}+1}$,
which is the maximal value it can reach in this case and it is
nonzero, so two atoms will be entangled even if they are separable
initially. Here, although both $A$ and $B$ are modified by a
grey-body factor $f(\omega,r)$, the modification cancels when we
evaluate $R$. So, $R$ is the same as what we get in the case of a
thermal bath. This result shows that, both in the vicinity of the
horizon and at infinity, the impact of the black hole to the
equilibrium entanglement of the system is the same as that of a
thermal bath at temperature $T=T_H/\sqrt{g_{00}}$ in a flat
space-time. As $r\rightarrow \infty$, the effective temperature
becomes the Hawking temperature, since the acceleration needed to
maintain the two-atom system at a fixed distance vanishes, and the
temperature is purely due to the thermal bath the black hole
immersed in.

 \subsection{The Boulware vacuum}

For the Boulware vacuum, the Wightman function is given by~\cite{wightman1,wightman2,wightman3}
 \begin{equation}
 G^+(x,x')=\sum_{lm}\int_{0}^{\infty}\frac{e^{-i\omega \Delta{t}}}{4\pi\omega}|\,Y_{lm}(\theta,\phi)\,|^2\big[|\,\overrightarrow{R_l}(\omega,r)\,|^2
+|\,\overleftarrow{R_l}(\omega,r)\,|^2\big]d\omega\;.
 \end{equation}
The Fourier transform with respect to the proper time is
\begin{eqnarray}
\mathcal{G}({\lambda})&=&\int^{\infty}_{-\infty}e^{i\lambda{\tau}}{G^+}[x(\tau)]d\tau
=\sum_{ml}\frac{2l+1}{8\pi\lambda}\big[|\,\overrightarrow{R_l}({\lambda}\sqrt{g_{00}},r)\,|^2
+|\,\overleftarrow{R_l}({\lambda}\sqrt{g_{00}},r)\,|^2\big]\theta(\lambda)\;,
\label{gbf}
\end{eqnarray}
where $\theta(\lambda)$ is the step function. Plugging
Eq.~(\ref{gbf}) into Eq~(\ref{abc}), we have
\begin{eqnarray}\label{bab1}
A=B=\sum_{l=0}^{\infty}\frac{2l+1}{16\pi\omega}\big[\;|\,\overrightarrow{R_l}(\omega,r)\,|^2
+|\,\overleftarrow{R_l}(\omega,r)\,|^2\big]\;.
\end{eqnarray}
So
\begin{equation} R=\frac{B}{A}=1\;,
\end{equation}
everywhere outside the event horizon, and the  concurrence of the
equilibrium state is
\begin{equation}
{\cal C}[\rho^\infty]={\cal C}[\rho^\infty]_{max}={1\over 2}\;
\end{equation}
 which is the same as that in a
Minkowski vacuum. No thermal radiation is present.

\section{Conclusion}

In summary, we have examined the asymptotic entanglement between two
mutually independent two-level atoms at a fixed radial distance
outside a Schwarzschild black hole  in the paradigm  of open quantum
systems. We treat the two-atom system  as an open quantum system in
a bath of fluctuating quantized massless scalar fields in vacuum and
have studied the Boulware, Unruh, and Hartle-Hawking vacua
respectively.

In the Hartle-Hawking  and the Unruh vacuum cases, the concurrence
attains a non-zero value less than 1/2 as long as the two-atom
system is not placed right at the event horizon, indicating that the
two atoms will turn out to be entangled even if they are initially
separable. For the Unruh vacuum case, the concurrence is the same as
if there were an outgoing thermal flux of radiation from the event
horizon, which is backscattered by the curvature of the space-time.
For the Hartle-Hawking vacuum case, the concurrence behaves as if
the atom were in a thermal bath of radiation at a proper temperature
which reduces to the Hawking temperature in the spatial asymptotic
region. Remarkably, only in the Unruh vacuum, will  the asymptotic
entanglement be affected by the backscattering of the thermal
radiation off the space-time curvature. The effect of the back
scatterings on the asymptotic entanglement cancels in the
Hartle-Hawking vacuum case.

\begin{acknowledgments}
We would like to thank Wenting Zhou for valuable discussions. This
work was supported in part by the National Natural Science
Foundation of China under Grants No. 11075083 and No. 10935013; the
Zhejiang Provincial Natural Science Foundation of China under Grant
No. Z6100077; the National Basic Research Program of China under
Grant No. 2010CB832803; the PCSIRT under Grant No. IRT0964; the
Hunan Provincial Natural Science Foundation of China under Grant No.
11JJ7001, and the Program for the Key Discipline in Hunan Province.
\end{acknowledgments}


\begin{thebibliography}{00}

\bibitem{hk}
S. Hawking, Nature, {\bf 248}, 30 (1974);
S. Hawking, Commun. Math. Phys. {\bf 43}, 199 (1975).

\bibitem{Gibbons}
G. Gibbons and S. Hawking, Phys.\ Rev.\ D {\bf 15}, 2752 (1977).

\bibitem{Parikh}
M. Parikh and F. Wilczek, Phys. Rev. Lett. {\bf 85}, 5042 (2000).

\bibitem{SC}
A.~Strominger and C.~Vafa, Phys. Lett. {\bf B 379}, 99(1996);
A.~Peet, hep-th/0008241.

\bibitem{Robinson}
S.~P.~Robinson and F.~Wilczek, Phys. Rev. Lett. {\bf 95}, 011303 (2005);
S.~Iso, H.~Umetsu, and F.~Wilczek, Phys. Rev. Lett. {\bf 96}, 151302 (2006).

\bibitem{hkur}
S. Deser and O. Levin, Phys. Rev. D {\bf 59}, 064004 (1999).

\bibitem{yu1}
H. W. Yu and W. Zhou, Phys. Rev. D {\bf 76}, 027503 (2007); {\bf
76}, 044023 (2007).

\bibitem{yu2}
W. Zhou and H.W. Yu, Phys. Rev. D {\bf 82}, 104030 (2010).

\bibitem{yu3}
H. W. Yu and J. Zhang, Phys. Rev. D {\bf 77}, 024031 (2007).

\bibitem{ZY}
H. W. Yu and J. Zhang, {\it Proceedings of the Ninth Asia-Pacific
International Conference on Gravitation and Astrophysics, Wuhan,
China, 2009}, edited by J. Luo et al (World Scientific Publishing,
Singpore, 2010), p. 319.

\bibitem{Benatti1}
F. Benatti and  R. Floreanini , Phys. Rev. A {\bf 70}, 012112
(2004).

\bibitem{yu4}
H. W. Yu, Phys. Rev. Lett. {\bf 106}, 061101 (2011).

\bibitem{pr1}
D. Braun, Phys. Rev. Lett. {\bf 89}, 277901 (2002).

\bibitem{pr2}
M. S. Kim, J. Lee, D. Ahn and P. L. Knight, Phys. Rev. A {\bf 65},
 040101(R)  (2002).

\bibitem{pr3}
L. Jakobczyk, J. Phys. A {\bf 35}, 6383 (2002).

\bibitem{pr4}
S. Schneider and G. J. Milburn, Phys. Rev. A {\bf 65}, 042107 (2002).

\bibitem{pr5}
F. Benatti, R. Floreanini and M. Piani, Phys. Rev. Lett. {\bf 91},
 070402  (2003).

\bibitem{pr6}
A. M. Basharov, J. Exp. Theor. Phys. {\bf 94}, 1070 (2002).

\bibitem{Unruh}
W. G. Unruh, Phys. Rev. D {\bf 14}, 870 (1976).

\bibitem{Hartle-Hawking}
J.~Hartle and S.~Hawking, Phys. Rev. D {\bf 13}, 2188 (1976).

\bibitem{Boulware}
D. G. Boulware, Phys. Rev. D {\bf 11}, 1404 (1975).

\bibitem{Benatti2}
F. Benatti and  R. Floreanini,  J. Opt. B {\bf 7}, S429 (2005).

\bibitem{Lindblad}
V. Gorini, A. Kossakowski, and E. C. G. Surdarshan, J. Math. Phys.
{\bf 17}, 821 (1976); G. Lindblad, Commun. Math. Phys. {\bf 48}, 119
(1976).

\bibitem{wightman1}
B. S. DeWitt, Phys. Rep. {\bf19}, 295 (1975).
\bibitem{wightman2}
S. M. Christensen and S. A. Fulling, Phys. Rev. D {\bf 15}, 2088
(1977).
\bibitem{wightman3}
P. Candelas, Phys. Rev. D {\bf 21}, 2185 (1980).

\bibitem{zhang}
J. Zhang and H.W. Yu, Phys. Rev. A {\bf75}, 012101 (2007); J. Zhang
and H. W. Yu, Phys. Rev. D {\bf75}, 104014 (2007).


\end{thebibliography}
\end{document}